# Spiral defect chaos in a model of Rayleigh-Bénard convection

Hao-wen Xi, J.D. Gunton

Department of Physics

Lehigh University

Bethlehem, Pennsylvania 18015,

and

Jorge Viñals

Supercomputer Computations Research Institute, B-186

Florida State University, Tallahassee FL 32306-4052

and

Department of Chemical Engineering, B-203

FAMU/FSU College of Engineering

Tallahassee, Florida 32316-2185.

October 24, 2018

# Abstract

A numerical solution of a generalized Swift-Hohenberg equation in two dimensions reveals the existence of a spatio-temporal chaotic state comprised of a large number of rotating spirals. This state is observed for a reduced Rayleigh number $\epsilon = 0.25$. The power spectrum of the state is isotropic, and the spatial correlation function decays exponentially, with an estimated decay length $\xi \approx 2.5\lambda_c$, where $\lambda_c$ is the critical wavelength near the onset of convection. Our study suggests that this spiral defect state occurs for low Prandtl numbers and large aspect ratios.

The spatio-temporal chaotic behavior of spatially extended, dissipative systems has been intensively studied in recent years [1]. The transition to spatio-temporal chaos has been observed in various physical systems such as Rayleigh-Bénard convection [2], optical instabilities [3], flames [4], and chemical reactions [5]. Spatio-temporal chaos manifests itself through a breakdown of global spatial coherence. However, a macroscopic *coherence length* –a length scale below which the pattern appears coherent– may still be observed. This form of chaos is often referred to as *weak turbulence.* Recently, a chaotic spiral defect pattern (which we will refer to loosely as spiral chaos below) has been observed in Rayleigh-Bénard convection in a non-Boussinesq fluid ($CO_2$ gas) [6], for a large aspect ratio and at moderate Rayleigh numbers. In previous experiments on convection in $CO_2$ gas [7], the spontaneous formation of a stable rotating spiral pattern was observed at a lower Rayleigh number. The chaotic state observed more recently is comprised of a large number of rotating spirals. Spirals nucleate, interact and annihilate yielding a macroscopically disordered pattern. In this paper, we show that a generalized Swift-Hohenberg equation which includes a quadratic nonlinearity and coupling to mean flow can account for the formation of this chaotic spiral pattern. We find that chaotic spiral patterns are spontaneously formed during the transition from the conduction state to rolls, in agreement with the experimental observations.

It is important to distinguish from the outset among the chaotic dynamics of dissipative systems that involve a small number of degrees of freedom, the chaotic spatio-temporal behavior of systems with many degrees of freedom, and fully developed turbulence. We follow Hohenberg and Shraiman [8], who defined three characteristic length scales: the dissipation length $l_D$, the excitation length $l_E$ and the correlation length $\xi$. The ratios of these lengths with respect to each other and with respect to the characteristic system length $L$ are used to characterize the chaotic state. The dissipation length $l_D$ is the characteristic length at which energy is dissipated. The excitation length $l_E$ is the characteristic length in which energy is injected into the system. Thus, for example, near the onset of Rayleigh-Bénard convection in an in-finite system, $l_D \approx k_c^{-1}$, where $k_c$ is the critical wavenumber. The excitation length $l_E \approx d \approx k_c^{-1}$, where $d$ is the thickness of the fluid layer. For a small aspect ratio and for moderate values of the reduced Rayleigh number $R/R_c$ (where $R_c$ is the critical Rayleigh number), we have $L \approx l_D \approx l_E$ so that only a few modes are excited. Hence the dynamics of such systems can then be explained by considering only a small number of modes. At large reduced Rayleigh numbers $R/R_c >> 1$, one has $l_D << l_E$, which is the fully developed turbulent regime. We focus here on the case of moderate $R/R_c$, but a large system $L >> l_E$. If $\xi > L$, we would be still effectively dealing with a small system, coherent in space but that may be either regular or chaotic in time. When $\xi << L$, on the other hand, the dynamical behavior is incoherent in space, and the system may be chaotic both in time and space. This regime that occurs for $L >> l_E \approx l_D \approx d$ at moderate $R/R_c$, is the regime of spatio-temporal chaos or weak turbulence, and is characterized by a dominant macroscopic coherence length $\xi$.

We model convection in a non-Boussinesq fluid by a two-dimensional generalized Swift-Hohenberg model [9, 10], defined by Eqs. (1)-(4) below, which we solve by numerical integration. The Swift-Hohenberg equation and various generalizations of it have proven to be quite successful in explaining many of the features of convective flow in fluids, particularly near onset [11, 12, 13, 14]. As we show in this paper, the same holds true for the formation of spiral chaos in non-Boussinesq fluids. Our model is defined in dimensionless units by

$$\frac{\partial \psi(\vec{r},t)}{\partial t} + g_m \vec{U} \cdot \nabla \psi = \left[ \epsilon' - \left( \nabla^2 + 1 \right)^2 \right] \psi - g_2 \psi^2 - \psi^3, \tag{1}$$

$$\left[ \frac{\partial}{\partial t} - Pr(\nabla^2 - c^2) \right] \nabla^2 \zeta = \left[ \nabla(\nabla^2 \psi) \times \nabla \psi \right] \cdot \hat{e}_z, \tag{2}$$

where $\vec{U}$ is the mean flow velocity,

$$\vec{U} = (\partial_y \zeta) \hat{e}_x - (\partial_x \zeta) \hat{e}_y. \tag{3}$$

The boundary conditions are,

$$\psi|_B = \hat{n} \cdot \nabla \psi|_B = \zeta|_B = \hat{n} \cdot \nabla \zeta|_B = 0, \tag{4}$$

where $\hat{n}$ is the unit normal to the boundary of the domain of integration, $B$. Equation (1) with $g_2 = g_m = 0$ reduces to the Swift-Hohenberg (SH) equation. The scalar order parameter $\psi(\vec{r},t)$ is related to the fluid temperature in the mid-plane of the convective cell, and $\zeta(\vec{r},t)$ is the vertical vorticity potential. Mean flow arises when the vertical vorticity is driven by roll curvature and amplitude modulations. Coupling to mean flow has been shown to play a key role, for example, in the onset of weak turbulence in Boussinesq fluids [10, 15, 16]. The quantity $\epsilon'$ is the scaled control parameter, $\epsilon' = (\frac{4}{k_c^2 \xi_0^2})\epsilon$, where $\epsilon = (\frac{R}{R_c} - 1)$ is the reduced Rayleigh number. Here $R$ is the Rayleigh number, $R_c$ is the critical Rayleigh number for an infinite system, $k_c$ is the critical wave number, $\xi_0$ is a characteristic length scale, $Pr$ is the Prandtl number, and $c^2$ is an unknown constant.

The values of the parameters that enter the equation have be chosen in the range appropriate for earlier experiments of Bodenschatz et al on $CO_2$ [7]. In order to estimate them in terms of experimentally measurable quantities, we have derived a three mode amplitude equation from the generalized Swift-Hohenberg equation [14]. From the experiments described in [7], we have estimated $g_2 \approx 0.35$ and $g_m \approx 50$. The value of $\epsilon'$ used in the numerical solution is 0.7, which is related to the experimental value $\epsilon$ in Ref. [7] by $\epsilon = 0.3594\epsilon'$=0.2516. We have chosen $c^2 = 2$ to simulate approximately the experimental rigid-rigid boundary condition. (Note that $c^2 = 0$ corresponds to a free-free boundary condition.) We note that for these values of the parameters, the classical work of Busse [17] on the stability of various convective states in a infinite fluid predicts that the stable pattern is a set of parallel rolls.

In the numerical calculations we consider a circular cell of radius $R = 32\pi$, which corresponds to an aspect ratio $\Gamma = R/\pi = 32$. A square grid with $N^2$ nodes has been used with spacing $\Delta x = \Delta y = 64\pi/N$, and $N = 512$. We approximate the boundary conditions on $\psi$ by taking $\psi(\vec{r},t) = 0$ for $\|\vec{r}\| \geq R$, where $\vec{r}$ is the location of a node with respect to the center of the domain of integration. In order to study the formation of the chaotic spiral pattern from the conducting state, we choose as initial condition $\psi(\vec{r}, t = 0)$ a random variable, Gaussianly distributed with zero mean and a variance 0.001.

Our main result is that this model exhibits a spatio-temporal chaotic spiral pattern which is remarkably similar to that observed experimentally [6], (there the system was quenched from the conduction state as well). Figure 1 shows a typical configuration that we have obtained exhibiting spiral chaos. Dark regions correspond to hot rising fluid and white regions to cold descending fluid. Initially, after the control parameter is quenched into the regime in which spiral chaos arises, the randomness in the initial configuration is rapidly lost. On a time scale $t \approx 600$, the system self-organizes into a regular structure comprising locally rotating spirals that fill the entire cell.

The two dimensional power spectrum $< P(\vec{k}) > = \left\langle |\psi(\vec{k})\psi(-\vec{k})|^2 \right\rangle$ is shown in Fig. 2(a), where $\psi(\vec{k})$ is the spatial Fourier transform of $\psi(\vec{r}, t)$ and $<>$ denotes a time average over one horizontal diffusion time [19]. The most interesting feature is that the intensity of the spectrum appears to be isotropic. Figure 2(b) shows the circularly averaged power spectrum $P(k)$. We see that $P(k)$ is broad, skewed and peaked at a wavenumber $k_{max} < k_c = 1$. In order to estimate the width of $P(k)$ we have fit it to a function of the form $A/\left(1 + \xi^4 \left(k^2 - k_0^2\right)^2\right)$. The fit obtained is shown in Fig. 2(b) as well, and corresponds to $A \approx 7.0, \xi \approx 2.4$ and $k_0 \approx 0.8$. The function $P(\vec{k})$ is more sensitive to position correlations than orientation correlations in the pattern. The dependence of a correlation function that would measure orientation correlations on distance need not be the same as that of $C(\vec{r})$ [18].

We finally show in Fig. 3 the field $\zeta$ that corresponds to the configuration shown in Fig. 1. White and dark regions correspond to clockwise and counterclockwise rotation respectively. The Fourier transform of this field shows, as expected, that it is peaked around $k = 0$ and is isotropic.

Our numerical investigation indicates that both large scale mean flow and large aspect ratio play a crucial role in the spontaneous formation of a spiral chaos state.

In the absence of the mean flow field, we do not observe it. To study how the spiral chaos state depends on the size of the system, a run was conducted in a small cell of aspect ratio 16. We observed a globally ordered pattern consisting of one two armed spiral, which demonstrates that a large aspect ratio is critical to the existence of spiral chaos. We have also studied the role of non-Boussinesq effects on the formation of the spiral chaos state. If the non-Boussinesq coupling constant $g_2 = 0$, and we start with the same random initial condition and use the same parameters ($\epsilon' = 0.7, g_m = 50, Pr = 1.0$ and $c^2 = 2.0$), we observe a similar spiral chaos state, demonstrating that the term that models non-Boussinesq effects is not necessary for spiral chaos. We have also carried out additional calculations to study the effect of the Prandtl number. Starting with exactly the same initial conditions and parameters as in Fig.1, but with a larger value of Prandtl number ($Pr = 6$), no spiral chaos pattern is observed, showing that low Prandtl number is essential for spiral chaos. The resulting pattern has a labyrinthine aspect. Finally, in order to study how generic the spiral chaos state is, we have studied the case in which the system was ramped slowly from the conduction state to the convective state. We used exactly the same initial conditions and the same parameters as in Fig .1, but increased $\epsilon$ very slowly up to $\epsilon = 0.7$: $\epsilon = 10^{-3}t$ for $0 < t < 700$, and $\epsilon = 0.7$ for $t > 700$. A similar spiral chaos is observed to form as time increase, although rolls tend to persist near the cell boundary. These results suggest that only low Prandtl number and large aspect ratio are relevant for the existence of spiral chaos.

# Acknowledgments

We wish to thank E. Bodenschatz, G. Ahlers and D. Cannell for suggesting the numerical investigation of the generalized Swift-Hohenberg equation, and them S. Morris, H.S. Greenside and P.C. Hohenberg for many stimulating conversations and comments. This work was supported in part by the National Science Foundation under Grant No. DMR-9100245. This work is also supported in part by the Supercomputer Computations Research Institute, which is partially funded by the U.S. Department of Energy contract No. DE-FC05-85ER25000. The calculations reported here have been carried out on the Cray Y-MP at the Pittsburgh Supercomputing Center.

**Figure captions**

Figure 1. Typical configuration of a spiral chaos pattern. The field $\psi$ is shown. Dark regions correspond to $\psi > 0$ and light regions to $\psi < 0$. The configuration shown has evolved from random initial conditions in a cylindrical cell with aspect ratio $\Gamma = 32$. The values of the parameters used are $g_2 = 0.35$, $g_m = 50$ and $c^2 = 2$. The parameter $\epsilon$ is quenched from 0 to 0.7 in the calculation. The configuration shown is at $t = 900 \approx t_h$.

Figure 2. (a) Power spectrum in $\vec{k}$-space. The intensity of the spectrum is approximately isotropic. (b) Circularly averaged power spectrum. The wavenumber $k$ is in units of $k_c$, where $k_c = 1$ is the critical wavenumber. The solid line is the fit discussed in the text.

Figure 3. Vorticity potential field $\zeta$ corresponding to the configuration shown in Fig. 1. White and dark regions correspond to clockwise and counterclockwise rotation, respectively.